\documentclass[12pt,preprint]{aastex}    
\shorttitle{Initial Helium content of globular clusters}       
\shortauthors{Cassisi, S., Salaris, M. \& Irwin, A. W.}       
       
\begin{document}       
       
\title{The initial Helium content     
of Galactic Globular Cluster stars from the $R$-parameter:    
comparison with the CMB constraint}       
    
\author{Santi Cassisi}       
\affil{INAF - Osservatorio Astronomico di Collurania,       
 via M. Maggini, 64100 Teramo, Italy; cassisi@te.astro.it}     
\affil{Max-Planck-Institut f\"ur Astrophysik,    
Karl-Schwarzschild-Strasse 1, 85741 Garching, Germany}    
       
\author{Maurizio Salaris}       
\affil{Astrophysics Research Institute, Liverpool John Moores        
       University, Twelve Quays House, Egerton Wharf, Birkenhead CH41        
       1LD, United Kingdom; ms@astro.livjm.ac.uk}       
\affil{Max-Planck-Institut f\"ur Astrophysik,    
Karl-Schwarzschild-Strasse 1, 85741 Garching, Germany}    
    
\author{Alan W. Irwin}       
\affil{Department of Physics and Astronomy, University of Victoria,     
P.O. Box 3055, Victoria, British Columbia, Canada, V8W 3P6;    
irwin@uvastro.phys.uvic.ca}

\received{}       
\accepted{}

\pagebreak        
\begin{abstract}       
    
Recent precise determinations of the primordial He-abundance     
($Y_p$) from cosmic microwave background (CMB) analyses and   
cosmological    
nucleosynthesis computations,    
provide $Y_p$=0.248$\pm$0.001. On the other hand,    
recent works on the initial He-abundance of Galactic globular     
cluster (GGC) stars, making use of the    
$R$ parameter as He-indicator,     
have consistently obtained $Y_{GGC}\sim$0.20.    
    
In light of this serious discrepancy that casts doubt on    
the adequacy of low mass He-burning stellar models,     
we have rederived the initial He-abundance for stars in two     
large samples of GGCs, by employing     
theoretical models computed using new and    
more accurate determinations of the Equation of State     
for the stellar matter, and of the uncertain    
$^{12}$C$(\alpha,\gamma)^{16}$O reaction rate.    
Our models include semiconvection during the central convective    
He-burning phase, while the breathing pulses are    
suppressed, in agreement with the observational constraints coming    
from the measurements of the $R_2$ parameter in a sample of clusters.    
    
By taking into account the observational errors    
on the individual $R$-parameter values, as well as    
uncertainties in the GGC [Fe/H] scale, treatment of convection and    
$^{12}$C$(\alpha,\gamma)^{16}$O reaction rate,    
we have obtained, respectively, a mean $Y_{GGC}$=0.243$\pm$0.006    
and $Y_{GGC}$=0.244$\pm$0.006 for the two studied GGC samples.     
These estimates are now fully consistent with $Y_p$ obtained from     
CMB studies. Moreover, the trend of the individual     
He-abundances with respect to [Fe/H] is    
consistent with no appreciable He-enrichment along the GGC     
metallicity range.     
    
\end{abstract}       
       
\keywords{cosmic microwave background --     
globular clusters: general -- stars: abundances --    
stars: evolution -- stars: horizontal branch}       
       
\pagebreak        
\section{Introduction}       
       
Galactic Globular Cluster (GGC) stars are the oldest objects    
in the Galaxy, and their initial He abundance ($Y_{GGC}$, where $Y$    
denotes the mass fraction of He)    
is supposed to be approximately equal to the primordial He abundance    
($Y_p$) produced during the Big Bang Nucleosynthesis (BBN).    
    
The value of $Y_p$ derived from spectroscopy of low-metallicity,    
extragalactic H~II regions, appears to be still subject to systematic    
uncertainties (see, e.g., the discussion in Bono et a.~2002 and    
references therein); as an example, Olive, Steigman \& Skillman~(1997)    
have determined $Y_p$=0.234$\pm$0.002, while Izotov \& Thuan~(1998)    
obtained $Y_p$=0.244$\pm$0.002, consistent with earlier findings by    
Kunth \& Sargent~(1983).    
    
On the other hand, recent determinations of the cosmological baryonic    
matter density ($\Omega_b$) from the cosmic microwave background (CMB) power spectrum     
obtained by the BOOMERANG, DASI and MAXIMA experiments    
provide consistently (e.g. Pryke et al.~2002,     
\"Odman et al.~2002; Sievers et al.~2002)     
a value $\Omega_b h^2$=0.022$\pm$0.003    
($h$ is the Hubble constant in units of 100 $Km Mpc^{-1} s^{-1}$).    
This baryon density coupled with standard BBN calculations    
(Burles, Nollett \& Turner~2001) provides $Y_p$=0.248$\pm$0.001;    
Such an independent determination of $Y_p$ is close to the    
spectroscopic determination by Izotov \& Thuan~(1998).    
     
As for the value of $Y_{GGC}$, empirical estimates    
are necessarily indirect, since    
He-lines are not detectable in GGC star spectra,    
apart from the case of hot Horizontal Branch (HB) objects, whose    
atmospheres are however affected by gravitational settling and    
radiative levitation, which strongly alter the initial chemical     
stratification (see, e.g., Michaud, Vauclair \& Vauclair~1983;     
Moehler et al.~1999).     
$Y_{GGC}$ estimates make use of results from stellar    
evolution, taking advantage of the fact that    
the evolution of low mass Population~II stars is    
affected by their initial He-content. More in detail, the so-called    
$R$-parameter (Iben~1968), defined as the number ratio of HB     
stars to Red Giant (RGB) stars brighter than the HB level     
($R=N_{HB}/N_{RGB}$), can be employed    
in order to determine $Y_{GGC}$ and therefore $Y_p$. The basic idea    
behind the use of this parameter as He indicator is that, at a given    
metallicity, a higher initial He-content implies a brighter HB and,    
in turn, a lower value of $N_{RGB}$ ($N_{HB}$ is only slightly affected),     
with the consequent increase of $R$.    
Other parameters derived from stellar evolution can     
also be employed (see, e.g., the discussion    
in Sandquist~2000 and Zoccali et al.~2000), but they are better suited to    
determine relative He-abundances than absolute ones.    
    
Following earlier analyses by Buzzoni et al.~(1983) and Caputo,    
Martinez Roger \& Paez~(1987),     
Sandquist~(2000 -- S00) and Zoccali et al.~(2000 - Z00) have recently    
estimated $Y_{GGC}$ by measuring the value of $R$ in large samples of    
GGCs (43 objects taken from various sources, in case of the S00 paper;     
26 objects with homogeneous $HST$ photometry for the Z00 paper), and compared    
the observational values with results from stellar evolution models.    
In both cases a value $Y_{GGC}\sim$0.20 was found, in severe    
disagreement with the CMB result and spectroscopic determinations for    
H~II regions. This large discrepancy between CMB and $R$-parameter    
results casts doubt on the    
ability of stellar models to accurately predict the evolutionary    
times of these crucial phases of stellar evolution. Reasons for    
this low He-content inferred from the $R$ parameter have been ascribed    
to the uncertainty of the $^{12}$C$(\alpha,\gamma)^{16}$O reaction    
rate, which is relevant    
during the late stages of central He-burning, but may in    
principle also be due, for example, to an improper treatment of the mixing in the     
central convective region of HB stars, which affect the HB    
evolutionary timescale, and hence $N_{HB}$ (see, e.g., Z00).    
    
In this paper we present a new determination of $Y_{GGC}$ from the     
$R$-parameter, employing in the model computation the most recent    
determination of the $^{12}$C$(\alpha,\gamma)^{16}$O reaction rate    
(Kunz et al.~2002) together with an improved equation of state (EOS)    
for the stellar matter (Irwin et al.~2002). We show that with these two     
improved physical inputs     
the discrepancy between CMB and $R$-parameter results     
almost completely disappears.    
This gives strong support to the accuracy of present HB stellar models and    
the adequacy of our convective core mixing treatment during the HB phase.     
In \S \ref{methods_sect} we briefly discuss the observational data and the    
theoretical stellar models, while in \S \ref{results_sect} and \S    
\ref{discussion_sect} the results are presented and discussed. Conclusions    
follow in \S \ref{conclusions_sect}.    
       
    
\section{Observational data and theoretical models \label{methods_sect}}       
    
We have determined $Y_{GGC}$ making use of     
our theoretical models and the observational     
databases presented by Z00 and S00. Z00 provide    
empirical $R$ values for    
26 GGCs spanning all the relevant metallicity range.     
The number of RGB stars is computed starting from the level      
of the observed V magnitude of the Zero Age HB     
(ZAHB -- lower envelope of the observed HB     
star distribution). This means that to compare the models with Z00    
data one has first to transform theoretical bolometric luminosities    
to $V$ magnitudes, and then determine the theoretical $R$ values    
following the definition of $R$ used by Z00.    
    
S00 provides $R$ values for 43 GGCs, but employs     
a slightly different definition of $R$, the same as Buzzoni et al.~(1983):     
the reference    
level for the RGB counts is the bolometric luminosity corresponding to the    
average $V$-magnitude of HB stars.    
In order to determine the level of the RGB corresponding to this    
bolometric luminosity S00 have applied to the observational data      
a relationship for the    
difference in bolometric corrections between HB and RGB stars    
(which is different from the one we applied to our models when using    
the Z00 definition for the $R$-parameter).    
    
We have compared our theoretical results with these two databases    
taking into account these two different definitions of $R$.    
The derived He-abundances will thus reflect the theoretical uncertainties     
related to the the different bolometric    
corrections employed in the two methods, and the different    
observational samples.    
    
In order to take into    
account current empirical uncertainties on the GGC metallicity scale we have    
used for the individual clusters [Fe/H] values given by    
Rutledge, Hesser \& Stetson.~(1997) on both the Carretta \&    
Gratton~(1997 -- CG97) and Zinn \& West~(1984 -- ZW84)     
scales (the internal accuracy    
of these [Fe/H] values is of the order of 0.10 dex). For clusters not    
listed by Rutledge et al.~(1997) we have used the original ZW84    
values transformed then to the CG97 scale using the conversion formula    
given by CG97.    
   
We have determined the existence of possible   
He-abundance vs [Fe/H] correlations by   
computing the corresponding correlation coefficient and evaluating its    
significance. In case of no correlation, we have determined $Y_{GGC}$ by means of a weighted   
mean of the individual values, with weights inversely proportional to   
the square of the individual errors. We notice that in Z00 the value    
of $Y_{GGC}$ was determined by simply considering the constant value of    
the He-abundance best fitting the individual datapoints,   
without taking into account individual errors.    
The existence of a significant spread in the individual cluster He-abundance   
has been studied by means of the statistical F-test.  
   
We computed new theoretical models and isochrones for     
Y=0.23, 0.245, 0.26, using -- like in Z00 analysis --     
the same code, input physics and bolometric corrections     
as in Cassisi \& Salaris~(1997), with the following modifications:    
    
\begin{itemize}    
\item We have updated the energy loss rates from plasma-neutrino    
processes using the most recent and accurate results provided     
by Haft, Raffelt \& Weiss~(1994).    
       
\item We have updated the nuclear reaction rates using the    
NACRE database (Angulo et al.~1999), with the exception of the     
$^{12}$C$(\alpha,\gamma)^{16}$O reaction. For this reaction we employ the     
more accurate recent determination by Kunz et al.~(2002), based on $\gamma$    
angular distribution measurements of $^{12}$C$(\alpha,\gamma)^{16}$O    
and a consistent ${\mathbf R}$-matrix analysis of the process. The     
claimed relative uncertainty of this new rate is half of the    
uncertainty quoted in previous determinations.    
    
\item An improved EOS whose code is made publicly available at    
\url{ftp://astroftp.phys.uvic.ca} 
\noindent 
\url{/pub/irwin/eos/code/eos\_demo\_fortran.tar.gz}    
under the GNU General Public License (GPL), has been used. A full description of this EOS    
is in preparation (Irwin et al.~2002) so we will only summarize its    
principal characteristics here.  The EOS is calculated using an    
equilibrium-constant approach to minimize the Helmholtz free-energy.  For    
realistic abundance mixtures, this approach greatly reduces the number of    
linear equations that must be solved per iteration so that the solution can    
be rapidly obtained.  This speed makes it practical to call the EOS directly    
from the stellar-interior code without introducing the errors associated    
with interpolating EOS tables (Cassisi \& Irwin 2002, see also Dorman,    
Irwin, \& Pedersen 1991).  The equilibrium-constant approach gives numerical    
solutions of high quality with thermodynamic consistency which is typically    
better than 1 part in $10^{11}$. The ``EOS1'' mode of the free-energy model    
that is used for the present calculations includes the following:    
arbitrarily relativistic and degenerate free electrons (Eggleton, Faulkner,    
\& Flannery 1973); a pressure-ionization occupation probability similar to    
that of Mihalas, Dappen, \& Hummer (1988); a Planck-Larkin occupation    
probability (Rogers 1986); the exchange effect for arbitrarily relativistic    
and degenerate electrons (Kovetz, Lamb, \& Van Horn 1972); and the Coulomb    
effect.  The Coulomb effect is treated with the Debye-H\"uckel approximation    
in the weak coupling limit and an approximation (Pols et    
al.~1995) of the multicomponent combination of    
the one-component plasma result (DeWitt, Slattery, \& Chabrier    
1996) in the strong-coupling limit.  A spline    
fit is used to interpolate between the weak and strong coupling limits. The    
size of the intermediate coupling region and the size of the interaction    
radii that characterize the pressure-ionization occupation probability are    
adjusted to fit the OPAL EOS tables distributed at    
\url{ftp://www-phys.llnl.gov/pub/opal/eos/}.    
    
\item We have explicitly taken into account    
the $\alpha$-enhanced chemical composition typical of Population~II stars,    
using the same initial metal mixture employed by    
Salaris \& Weiss~(1998), and their same opacity tables;    
the heavy element distribution has an average 
$\alpha$-enhancement equal to 0.4 dex.  
This is potentially important for the upper metallicity end of the GGCs,    
since in that regime the well known equivalence between low-mass scaled-solar     
and $\alpha$-enhanced models with the same total metallicity (Salaris,    
Chieffi \& Straniero~1993) is no longer valid    
(Salaris \& Weiss~1998, VandenBerg et al.~2000).
We just notice, in passing, that an average enhancement of 0.4 dex is in 
full agreement with abundance data from Halo field stars (e.g., the 
discussion in Salaris \& Weiss~1998), while the GGCs data compiled by 
Carney~(1996) seem to point out to an $\alpha$-enhancement slightly 
lower, of about 0.3 dex. Such a small difference -- if real -- does 
not introduce any serious bias in our final $Y_{GGC}$ estimates, 
because such a small difference in the $\alpha$-enhancement 
between clusters and models can be 
fully compensated (also at high metallicities) by a small rescaling of 
the Z-[Fe/H] relationship of the theoretical models, which introduces a 
systematic effect of less than 0.001 on the individual cluster 
He-abundance estimates.

\item We have accounted for the different evolutionary times characterizing   
the red and blue parts of the HB. Ordinarily, the theoretical values of $R$    
are computed -- as in Z00 -- by considering the HB evolutionary time of a    
star populating the middle of the RR~Lyrae instability strip    
(log$(T_{eff})$=3.85). This is strictly adequate only for those clusters    
with an HB populated at the RR~Lyrae instability strip and redward    
(increasing total stellar mass), since the HB evolutionary timescales are    
basically unchanged when moving from the instability strip towards the red    
(see the discussion in Z00). However, stars populating the HB blueward of    
the instability strip do show different evolutionary times, which increase    
for decreasing total stellar mass. At the bluest end of a typical blue HB    
the increase of the HB evolutionary time with respect to the RR~Lyrae strip    
counterpart can amount to about 20 \% (Z00). We will discuss in \S    
\ref{blue_sect}     
the implications for the derived He-abundance in GGCs with a blue    
HB.    
\end{itemize}

\section{The value of $Y_{GGC}$ from the $R$-parameter    
\label{results_sect}}     
    
In this section we present separately     
our determination of $Y_{GGC}$ using the Z00 and S00    
samples.    
    
\subsection{The Z00 sample}    
    
\begin{figure}       
\plotone{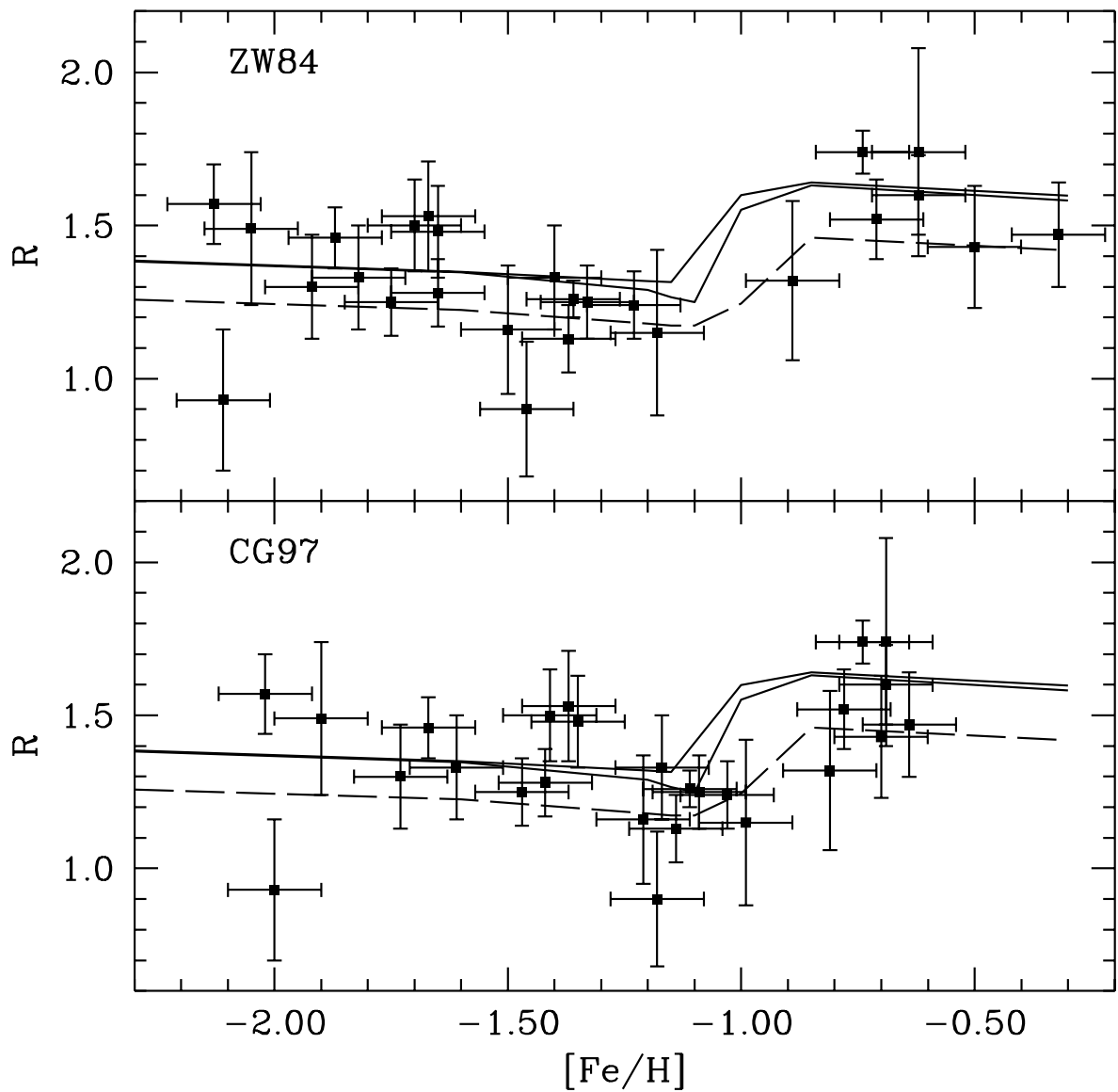}       
\caption{$R$-parameter versus [Fe/H] for the two adopted metallicity     
scales. Empirical data (filled squares) and individual errors      
are from Z00; errors on [Fe/H] have been set to 0.10 dex.     
Theoretical predictions for Y=0.245 and GGC ages of 11 and 13    
Gyr are shown as solid lines. The dashed line displays     
the theoretical prediction for Y=0.230 and a GGC age of 13 Gyr.\label{Rz00}}       
\end{figure}

Fig.~\ref{Rz00} displays the run of the empirical data by Z00 together with    
theoretical predictions for ages of 11 and 13 Gyr and $Y$=0.245 (solid    
lines), as a function of [Fe/H].     
At fixed age and $Y$ the theoretical value of $R$ is very    
slowly decreasing up to [Fe/H]$\sim -$1.15. Between [Fe/H]$\sim -$1.15    
and [Fe/H]$\sim -$0.85 $R$ increases steeply; this    
increase is due to the fact that the RGB bump, previously located at    
brightnesses larger than the ZAHB, moves below the ZAHB level with    
increasing metallicity, thus    
causing an abrupt decrease of the number of RGB stars brighter than    
the ZAHB (see, e.g., the discussion in Z00).    
At higher [Fe/H] values $R$    
is again only very mildly decreasing with increasing [Fe/H].    
It is also interesting to notice how the dependence of $R$ on age    
is restricted to the interval ranging from [Fe/H]$\sim -$1.15 to    
[Fe/H]$\sim -$0.85, which is exactly the metallicity range where the RGB bump    
crosses the ZAHB level. This is easily explained by the fact    
that the RGB bump luminosity does depend on the stellar age (e.g.,    
Cassisi \& Salaris~1997) while the    
ZAHB level is basically unaffected for ages typical of GGCs; in    
general higher ages shift the RGB bump location towards lower luminosities.    
    
In the same Fig.~\ref{Rz00} we also display the theoretical $R$ values    
for an age of 13 Gyr and $Y_{GGC}$=0.23, to show the sensitivity of $R$ to the    
model initial He-content. The average value of the derivative $\delta    
R$/$\delta Y$ is $\sim$10.     
    
We have estimated $Y_{GGC}$ and the associated 
1$\sigma$ dispersion of the individual abundances (the latter will be 
thoroughly discussed at the end of this section),
by first assuming an average age of 13 Gyr for the clusters (see, e.g., the analyses by    
VandenBerg~2000; Salaris \& Weiss~1998, 2002); the error on the    
individual cluster $Y$ values has been obtained from the quoted errors    
on the value of $R$. When considering all 26    
clusters together with the CG97 [Fe/H] scale,    
we obtained a weighted mean    
$Y_{GGC}$=0.240$\pm$0.003 (1$\sigma$ error). However, we    
found a clear correlation between $Y_{GGC}$ and [Fe/H] in the sense that    
the mean $Y_{GGC}$ obtained for clusters with [Fe/H] between    
$-$1.15 and $-$0.85 (the metallicity range influenced by the assumed    
cluster age) is $Y_{GGC}$=0.231$\pm$0.005, while for [Fe/H]$< -$1.15 and     
[Fe/H]$> -$0.85 we found, respectively, $Y_{GGC}$=0.247$\pm$0.005 and     
$Y_{GGC}$=0.244$\pm$0.003 (no correlation of the individual Y estimates with    
[Fe/H] has been found in these latter two metallicity ranges).    
It is evident that the mean values of Y determined for [Fe/H]$< -$1.15 and     
[Fe/H]$> -$0.85 are in good agreement while a substantial lower    
value is obtained for the clusters whose $R$ parameter is    
affected by the precise value of the age.    
We have therefore rederived the He-content with a different assumption    
about the cluster ages. Rosenberg et al.~(1999) and Salaris \& Weiss~(2002)    
have shown how clusters with [Fe/H] larger than $\sim -$1.2 (on the CG97    
metallicity scale) display a large age spread and are on average    
younger by $\approx$2 Gyr than the more metal poor clusters.    
We have therefore recomputed the values of Y assuming an age of 13 Gyr    
for the clusters with [Fe/H]$< -$1.2 and 11 Gyr for more metal rich    
ones.    
As expected, the mean Y values for [Fe/H]$< -$1.15 and     
[Fe/H]$> -$0.85 are unchanged, but this time, in the [Fe/H] range      
between $-$1.15 and $-$0.85, we obtain a mean $Y_{GGC}$=0.239$\pm$0.004    
which, within the 1$\sigma$ error bar, is in better agreement with    
the results at higher and lower metallicities. It is     
therefore important to notice that    
the precise individual cluster ages do matter when determining an    
accurate $Y_{GGC}$ value for clusters in this [Fe/H] range.    
    
We have repeated the previous analysis by employing the ZW84 [Fe/H]    
scale. Adopting an age of 13 Gyr for all clusters, we derive    
a mean $Y_{GGC}$=0.242$\pm$0.003 for the whole cluster    
sample, and we do not find any correlation between Y and [Fe/H].    
However, Salaris \& Weiss~(2002) have shown that, when considering the ZW84    
metallicity scale, clusters with [Fe/H]$> -$1.6 show a large age    
spread and are on average younger than the more metal poor ones (see    
also VandenBerg~2000).    
We therefore repeated the previous calculation considering an age of 13 Gyr    
when [Fe/H]$< -$1.6 and 11 Gyr at higher [Fe/H]; we obtain a mean     
$Y_{GGC}$=0.243$\pm$0.003, consistent with the value determined for a constant    
age of 13 Gyr. This result comes from the fact that, when using the    
ZW84 metallicities, there are no clusters    
populating the [Fe/H] range which is strongly affected by age.    
    
Another important question to be addressed is the significance of the dispersion    
of the individual cluster values around the mean $Y_{GGC}$. In    
particular, it is important to know if the observed 1$\sigma$ dispersion, of the    
order of 0.02, is entirely due to the error on the individual     
cluster estimates. To address this point we have applied the statistical F-test     
(see, e.g., an application to the case of GGC ages by     
Chaboyer et al.~1996; Salaris \& Weiss~1997, 2002) to our sample of He    
determinations. In case of the CG97 [Fe/H] scale we have restricted    
the analysis to the clusters within the metallicity range unaffected    
by age, so that an  age spread will not affect the observed    
He-abundance dispersion.     
For each individual cluster we have calculated     
a set of synthetic He-abundances by randomly generating -- using 
a Monte Carlo procedure -- 10000 abundance values, 
according to a Gaussian distribution with mean value equal to the observed 
mean $Y_{GGC}$, and $\sigma$ equal to the    
individual He-abundance error. This is repeated for all    
clusters in the selected sample and the 10000 values for each    
individual clusters are joined to produce an ``expected'' $Y_{GGC}$    
distribution for the entire cluster sample, 
on the assumption that the detected He-abundance spread is not 
intrinsic, but due just to the individual error bars.       
The F-test has been then applied in    
order to determine if this ``expected'' distribution has the same    
variance as the observed one. We state that a $Y_{GGC}$ range does exist    
if the probability that the two distributions have different variance    
is larger than 95\%. In case this condition is verified, the size of the true     
$Y_{GGC}$ range ($\sigma_Y$) can be estimated according to    
$\sigma_Y$=$(\sigma_{obs}^2-\sigma_{exp}^2)^{0.5}$,    
where $\sigma_{obs}$ and  $\sigma_{exp}$ are, respectively, the    
1$\sigma$ dispersion of the actual data and of the ``expected'' distribution.     
    
The result of this test applied to the Z00 sample with our two choices    
of the [Fe/H] scale indicates that the observed dispersion around the    
mean $Y_{GGC}$ is entirely due to the formal errors (the probability 
that the observed and the synthetic distributions have different variance is 
below 70\% in both cases) on the individual determinations; therefore no 
statistically significant spread in the individual He-abundances is found.    
    
\subsection{The S00 sample}    
    
\begin{figure}       
\plotone{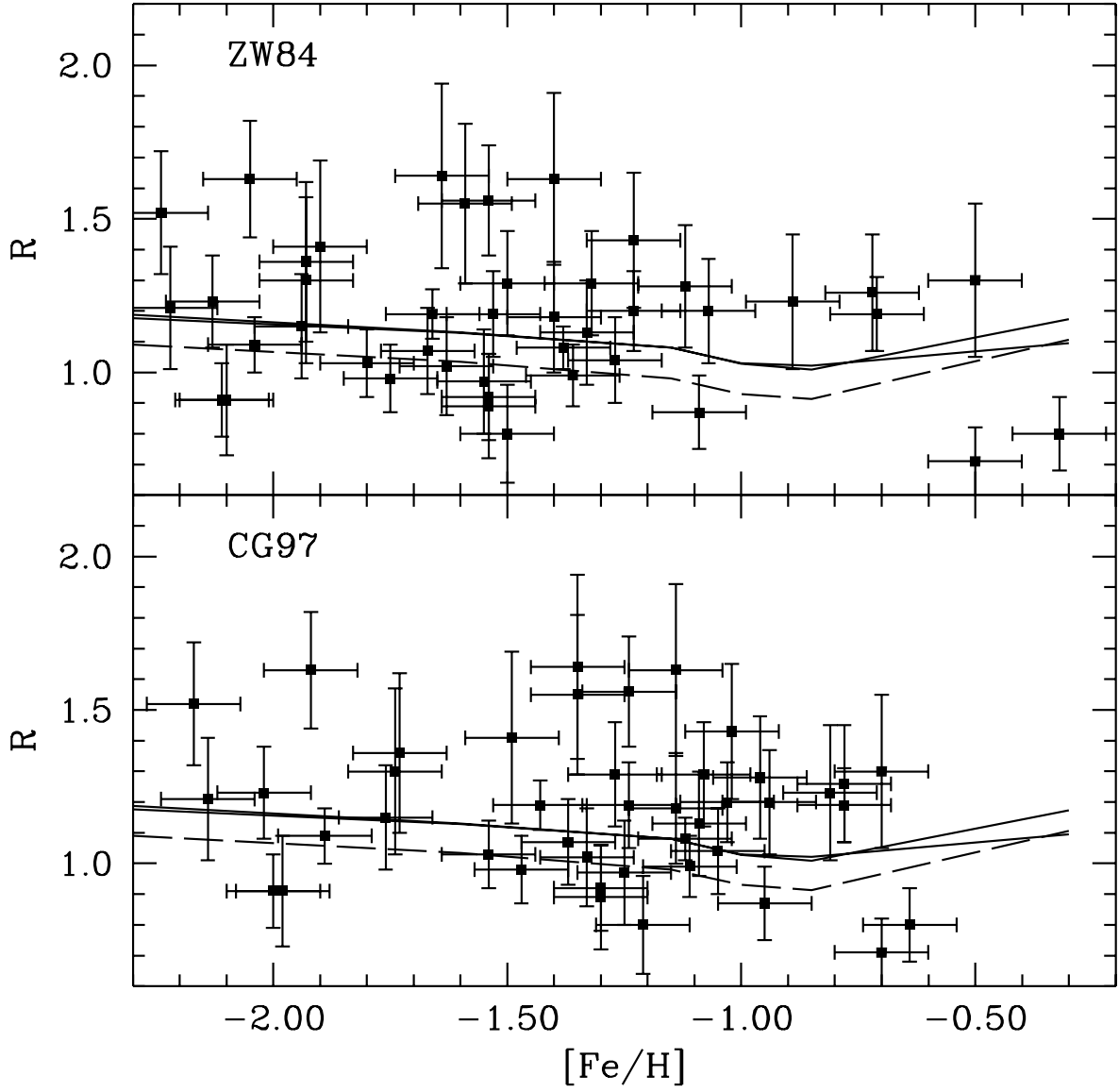}       
\caption{As in  Fig.~\ref{Rz00} but for the empirical data by S00.\label{Rs00}}       
\end{figure}

Fig.~\ref{Rs00} displays the run of the empirical data by S00 together with    
theoretical predictions for ages of 11 and 13 Gyr and $Y$=0.245 (solid    
lines), as a function of [Fe/H].     
At fixed age and $Y$ the theoretical value of $R$ is very    
slowly decreasing up to [Fe/H]$\sim -$0.85. At higher metallicities    
the value of $R$ increases, due again to     
the fact that the bolometric luminosity of the RGB bump crosses the     
reference HB bolometric luminosity. The shift to higher metallicities    
of this crossing region with respect to the $R$ definition previously used,    
arises from the fact that the bolometric luminosity of the RGB    
reference level corresponds to $V$ magnitudes fainter   
than the $V$ magnitude level of the HB.     
This implies that RGB bump stars are included into the    
determination of $R$ up to higher metallicities than the case of Z00 definition.     
This high metallicity region is also the only one affected by age (see    
discussion in \S 3.1); therefore    
the precise choice of the GGC ages does not affect at all the results when    
using the S00 definition of the $R$-parameter, since only very few    
clusters show these high values of [Fe/H] (see Fig.~\ref{Rs00}), and    
only on the ZW84 scale.    
    
By assuming t=13 Gyr for all GGC we find again a     
$Y_{GGC}$ distribution uncorrelated with [Fe/H].    
A mean value $Y_{GGC}$=0.246$\pm$0.005 is obtained when considering the CG97    
[Fe/H] scale, while $Y_{GGC}$=0.241$\pm$0.004 is derived when the ZW84    
metallicities are employed.    
As for the dispersion of the $Y_{GGC}$ values around    
these means, we obtain in both cases $\sigma_Y$=0.04; we have    
applied the F-test also in this case and derived that the dispersion    
can't be completely explained by the formal errors on the individual    
determinations (the probability that the variance 
of the He-abundance distribution in the observed 
sample and in the synthetic one are different is larger than 99\%), 
and has an intrinsic component equal to $\sigma_Y$=0.03    
(analogous conclusions were reached by S00).    
    
\subsection{Clusters with a blue HB \label{blue_sect}}    
    
All the $Y_{GGC}$ values given before have been obtained by considering the    
evolutionary time of HB stars in the instability strip when computing    
the theoretical value of $R$; this is also what has been done by Z00    
and S00.     
    
While this assumption is well founded in    
case of HBs populated at the strip and redward, it is less adequate in    
case of very blue HBs (see the discussion in Caputo et al.~1987 and    
Z00); this is particularly true when the location of the     
average mass populating the observed HB corresponds to $V$  
about 0.5-1.0 mag fainter than the instability strip level.    
This is due to the fact, as discussed in previous section, that the HB    
evolutionary times increase when one greatly reduces the total stellar    
mass with    
respect to the values attained at the instability strip.    
To correct for this possible systematic uncertainty caused by our assumption we    
have applied the following procedure.    
    
For both the Z00 and S00 samples we have identified those clusters    
whose HB is mainly populated at the blue side of the instability strip; among    
these clusters, through comparisons with our HB models, we have      
identified the objects whose average HB mass is located more than 0.7    
mag below the RR~Lyrae level. For these clusters, we have     
recomputed the theoretical $R$-values by taking as     
representative of the HB evolutionary lifetime the corresponding value for    
the average mass.    
There are only 6 clusters in the Z00 sample, and 8 clusters in the S00    
sample that satisfy this condition.     
    
\begin{figure}       
\plotone{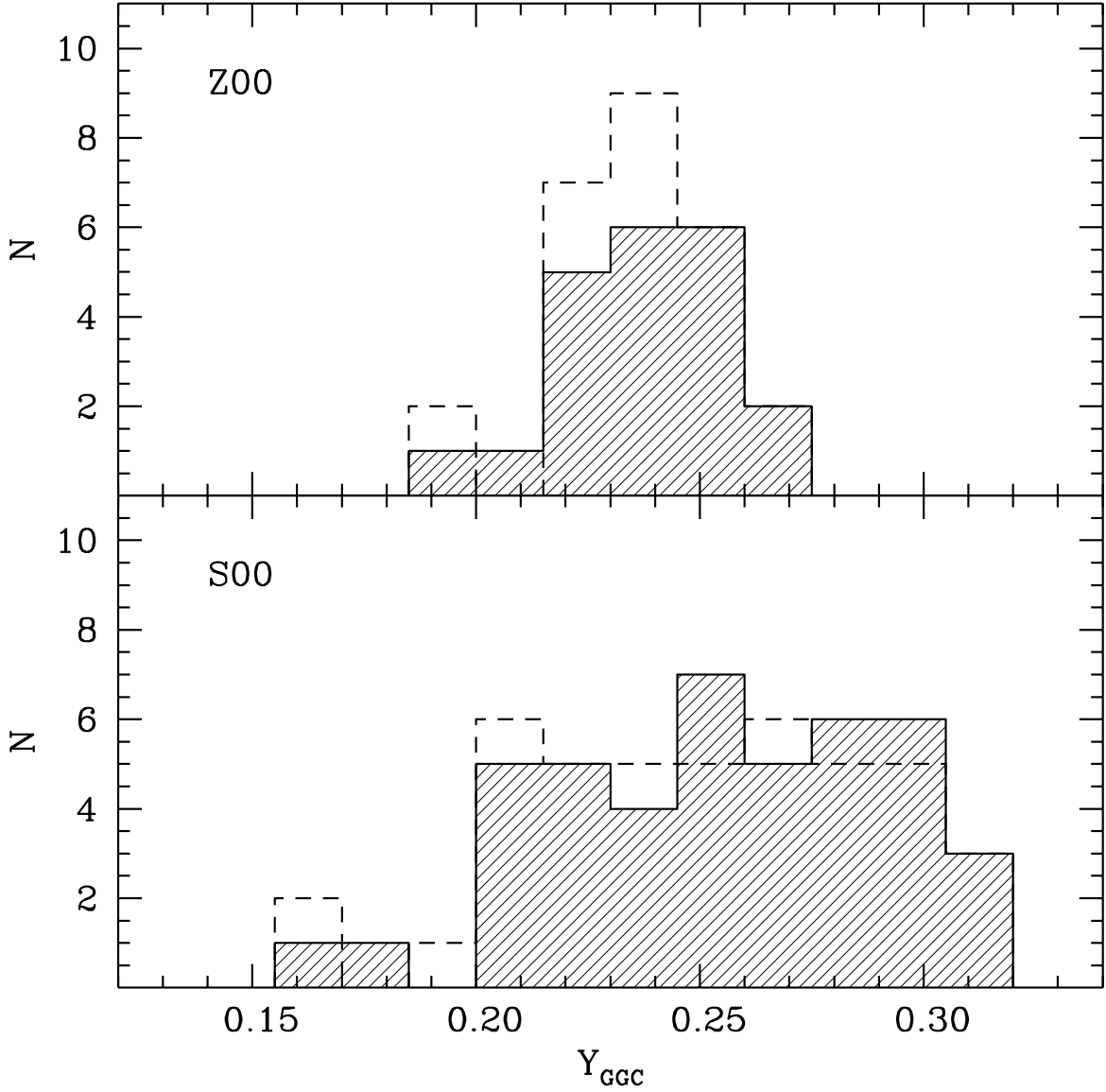}       
\caption{Histograms representing the distribution of the individual     
cluster He-abundances for the Z00 (upper panel) and S00 (lower panel) samples.    
Shaded histograms display the abundance distribution when the     
CG97 [Fe/H] scale is adopted; short-dashed lines represent the    
corresponding histograms for the ZW84 scale. In case of the Z00 data and the    
CG97 [Fe/H] scale we have included only clusters with [Fe/H]$< -$1.15 or    
[Fe/H]$> -$0.85, that is, the ranges unaffected by the precise    
choice of the GGC ages.\label{Rhist}}       
\end{figure}

When applying these corrected evolutionary times to the blue HB clusters    
we find that the $Y_{GGC}$ values obtained in the previous analysis    
are reduced by only 0.001-0.002.    
The size and significance of the $Y_{GGC}$    
spread, and the behaviour with the respect to [Fe/H], are unchanged with    
respect to the previous results.     
In Table~1 we summarize the $Y_{GGC}$ results, with and without the    
correction for the blue HB clusters. In case of the Z00 sample and the    
CG97 [Fe/H] scale we display the results for the metallicity range    
that is insensitive to the choice of the cluster ages.    
Fig.~\ref{Rhist} displays the distribution of the individual    
GGC He-abundance, for both samples and both choices of the [Fe/H]    
scale, taking into account the correction for the blue HB GGCs.    
The Z00 sample clearly has a significantly narrower abundance range than the    
S00 sample.    
       
\section{Discussion \label{discussion_sect}}    
    
In the previous section we found that the $R$ parameter provides    
values of $Y_{GGC}$ between $\sim$0.240 and $\sim$0.245,    
independent of [Fe/H]; the exact values are summarized in 
Table~1, together with the size of the intrinsic spread $\sigma_Y$ of 
the individual cluster He-abundances. 
     
The mean values of $Y_{GGC}$ deduced from the Z00 and S00 sample    
are in excellent agreement within the associated 1$\sigma$ error, in spite   
of the - in principle - different    
bolometric corrections applied to the data analysis and the different    
photometric samples employed.    
It is however important to mention the fact that the Z00 data do not    
provide any indication of a statistically significant     
spread of $Y_{GGC}$, while the opposite is    
true for the S00 data. 
One possible reason for this occurrence may be     
the inhomogeneity of the S00 sample, which is made of photometries    
taken with very different instruments and detectors (photographic,    
photoelectric and CCD photometries), reduced with different    
procedures in the course of the last 25    
years, and with possibly different methods to correct for incompleteness, 
as opposed to the homogeneously observed, reduced and analyzed    
$HST$ sample by Z00.    
 
Another possibility to explain this He-abundance spread is related to the 
existence of population gradients within the observed clusters, coupled with 
the fact that the $HST$ data employed by Z00 mainly sample regions of 
the clusters' cores, whereas the ground based photometries adopted by S00 
sample more external regions located at various distances from the 
cores. 
 
\begin{figure}       
\plotone{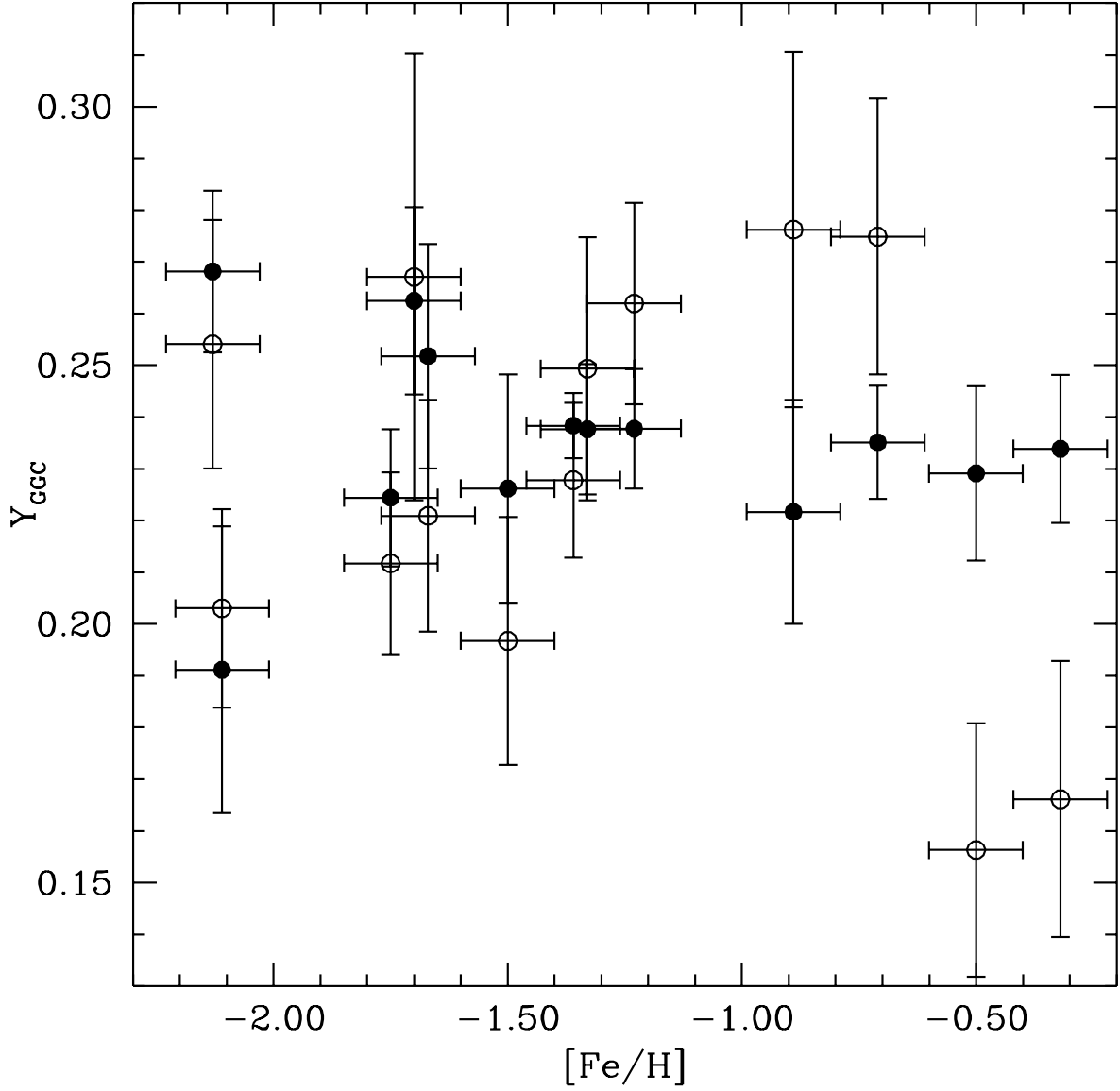}       
\caption{Helium abundance as a function of [Fe/H] (on the ZW84 scale)    
for 13 clusters in common between the Z00 (filled circles) and S00     
(open circles) samples.\label{Rcommon}}       
\end{figure}

We have also performed another test, by comparing the individual    
He-abundance for 13 clusters in common between the Z00 and S00    
samples. In Fig.~\ref{Rcommon} we display the abundances for these 13    
clusters derived from the Z00 and S00 data (we have chosen to use in    
this figure the ZW84 [Fe/H] scale, but this choice does not affect the    
result of the comparison), considering the corrections     
for the blue HB clusters. The Z00 data provide a mean     
$Y_{GGC}$=0.237$\pm$0.004, in very good agreement with the result from    
the whole sample displayed in Table~1 ($Y_{GGC}$=0.240$\pm$0.003); the    
dispersion around the mean is again due (as for the whole sample)    
only to the error on the individual determinations.    
In case of the S00 data for the same 13 clusters,     
the mean $Y_{GGC}$=0.224$\pm$0.006 is smaller than for the Z00    
data, and also significantly smaller than the mean value for the    
whole sample ($Y_{GGC}$=0.240$\pm$0.004); the dispersion around the    
mean $Y_{GGC}$ is larger than in the Z00 case.    
Therefore, whereas a    
somewhat random selected sizable cluster subsample     
(the 13 common clusters span all    
the relevant [Fe/H] range as well as show both red and blue HBs)    
show the same properties of the whole sample in case of the Z00 data,    
the opposite is true for the S00 data.    
This may lend some support to the idea that the significant dispersion    
of $Y_{GGC}$ for the whole S00 sample is due to some inhomogeneity intrinsic    
to the data used for determining the observed $R$ values. 
On the other hand, when the 4 most metal rich clusters 
([Fe/H]$> -$1) are excluded from 
the comparison shown in Fig.~\ref{Rcommon}, the dispersion of the S00 
data becomes comparable with the Z00 one, while the mean
He-abundance is similar to the value for the whole S00 sample.

This seems to point to some metallicity-related effect, which however does 
not explain the dispersion for the whole S00 sample. In fact, if we 
apply the F-test discussed in Sections 3.1 and 3.2 to the S00 sample 
without the clusters with [Fe/H]$> -$1, we still obtain a statistically 
significant dispersion of the individual He-abundances.
 
In spite of this difference regarding the spread in the cluster He-abundance   
for the Z00 and S00 samples,    
our results clearly indicate a mean value of    
$Y_{GGC}$ which is not in significant contradiction with the CMB    
constraint. This is very different from the conclusions reached    
by Z00 and S00 analyses, which derived an unrealistically     
low He-abundance, namely    
$Y_{GGC}\sim$0.20, completely inconsistent with the CMB constraint.    
When we redetermine $Y_{GGC}$  by using the same observational data and    
theoretical scenario adopted by Z00, but using the same weighted    
average method employed in our analysis and   
considering the metallicity range unaffected by the selected cluster age, we   
obtain $Y_{GGC}$=0.21, still largely incompatible with the CMB constraint.   
   
The new $^{12}$C$(\alpha,\gamma)^{16}$O reaction rate and the new EOS are the     
two physical ingredients that have strongly modified the theoretical $R$    
values with respect to the results by Z00, whose employed stellar models    
we have updated for this work. In particular, the recent estimate of    
the $^{12}$C$(\alpha,\gamma)^{16}$O reaction rate (Kunz et al.~2002)     
has reduced the HB    
evolutionary times (at a fixed core mass and envelope composition)   
by $\sim 7-8$\%, and the new EOS has further    
reduced the HB evolutionary times by $\sim$ 10\%. On the other hand,   
the new EOS also slightly reduces   
the value of the He-core mass at the He-flash   
for a given age, which has the effect of increasing by $\sim$2\% the HB   
evolutionary time; there is also a further     
increase by $\sim$4 \% for the value of   
$N_{RGB}$ because a larger portion of the RGB is considered in the   
evaluation of the $R$-parameter.   
These effects cause a total reduction of $R$ by $\sim$20 \% which, for a   
typical average observed value of $R$ (i.e., with the Z00 definition   
of $R$) equal to $\sim 1.4-1.5$, corresponds   
to an increase of the estimated $Y_{GGC}$ by about 0.03.   
    
The $^{12}$C$(\alpha,\gamma)^{16}$O reaction rate by Kunz et al.~(2002)    
has a relative uncertainty of about $\pm$30\%, which translates into a    
systematic uncertainty of about $\pm$0.008 around the values obtained    
in our analysis. It is also very interesting to notice at this point     
that Metcalfe \& Handler~(2002) find, from     
asteroseismology data for two local white dwarfs,    
central Oxygen abundances consistent with value obtained by using    
the $^{12}$C$(\alpha,\gamma)^{16}$O rate by Kunz et    
al.~(2002) during the progenitor He-burning phase.    
    
We believe the EOS calculations for the current set of stellar models do not    
contribute significant errors to the final results. The new EOS has been    
adjusted to fit the tabulated OPAL results.  The quality of the fit is    
quite good.  For example, the residuals for solar conditions are less    
than 0.06\% in the pressure, and this good agreement should also extend to    
all but the highest density portions of evolved models where there are some    
EOS uncertainties in the treatment of the Coulomb and electron exchange    
effects. However, the large variety of effects on the models caused by these    
non-ideal effects largely cancel each other so the calculated $R$ values are    
insensitive to these uncertainties.    
    
The Coulomb effect arises because the attractive Coulomb force between ions    
and electrons tend to correlate the two kinds of particles. The    
exchange effect arises because the total eigenfunction of electrons, which    
is antisymmetric with respect to electron exchange, anti-correlates  
the electrons with each other.  For fixed density and temperature, both the    
Coulomb correlation and the exchange anti-correlation reduce the amount of    
pressure required to confine the gas to its volume and also reduce the    
adiabatic gradient.      
    
To determine how Coulomb and exchange effects  
alter the theoretical R values, we did some test stellar-evolution 
calculations with and without the Coulomb or exchange effects for main 
sequence, RGB and HB phases. 
The calculated $R$ value is equal to $t_{HB}/t_{RGB}$, where $t_{HB}$ is  
the duration of the HB evolutionary phase, and $t_{RGB}$ is the duration  
of that part of the RG phase whose luminosity is greater than the luminosity  
of the HB. 
By analyzing the various numerical experiments, we found that although the 
evolutionary rate along the RGB is slightly affected by Coulomb and exchange 
effects, the quantity $t_{RGB}$ is not significantly changed as a consequence of 
the variation of the HB luminosity level which compensates the change in the RGB 
evolutionary rate. On the contrary, we found the following results for 
$t_{HB}$:
 
\begin{itemize}    
\item Coulomb and exchange effects for the precursor phase (i.e. the MS and the 
RGB) decrease the    
core mass by a small amount, but this reduction in fuel is more than    
compensated by the accompanying decrease in the helium burning luminosity so the total    
precursor effect for Coulomb and exchange is a 4\%     
increase in $t_{HB}$. The exchange effect accounts for about one sixth    
of this total.    
\item The Coulomb effect (which for weak coupling and full ionization is    
proportional to the cube of the atomic number of the element) is    
considerably enhanced for later stages of the HB phase because He-burning in    
the core substantially increases the abundance of C and O.    
\item Coulomb and exchange effects for the HB phase increase the convective core    
mass by roughly 5\%, but that is more than compensated by a helium burning   
luminosity increase of roughly twice as much.  Thus, the total HB effect for    
Coulomb and exchange is a 6\% decrease in $t_{HB}$.  The exchange effect    
accounts for about one third of this total.    
\item When one combines the opposite precursor and HB effects    
together, a further cancellation occurs so the total effect for Coulomb and    
exchange is only a 2\% decrease in $t_{HB}$.    
\item An alternative spline fit to the Coulomb effect (see the    
EOS description in \S \ref{methods_sect}) with a substantially enlarged    
range of intermediate coupling, changed the Coulomb results by about 10 per    
cent of their size.  This translates to a 0.2\% uncertainty in $t_{HB}$ and    
calculated $R$, and a negligible uncertainty in the derived $Y_{GGC}$ value.    
\end{itemize}    
    
Models of stellar interiors are sensitive to EOS interpolation errors    
(Dorman, Irwin, \& Pedersen 1991) so the most reliable calculational    
procedure is to eliminate EOS interpolation errors by calling the EOS code    
directly from the stellar interior code.  The present EOS is fast enough so    
that such direct use is practical on workstation type computers, but of    
course still substantially slower than calculations done with interpolated    
EOS tables.  Thus, in the interests of reducing the required computer time    
for the computations we interpolated tables of EOS results that were    
tabulated with the present EOS for the required ranges of pressure,    
temperature, $Y$, $X_C$, $X_N$, and $X_O$ for a fixed non-CNO metal    
abundance mix.  The adopted grid spacings are small enough in all    
coordinates so that the resulting $t_{HB}$ values gave excellent agreement    
with one test calculation using direct EOS results.    
   
As an additional test for the adequacy of our models and therefore of our    
inferred $Y_{GGC}$, we have also considered the so-called $R_2$ parameter,    
defined as the number ratio of Asymptotic Giant Branch (AGB) to HB stars    
(Caputo et al.~1989). This parameter is strongly sensitive to the extension    
of the convective cores during the HB phase, while it is fairly insensitive    
to the initial metal and He-abundance of the models, and the precise value    
of the age.  A test for our treatment of the convection in the HB stellar    
cores is of fundamental importance, since the extension of the convective    
core strongly affects the evolutionary time along the HB phase. An    
underestimate of the size of the HB convective cores would cause an    
underestimate of the HB evolutionary times, with a consequent spurious    
increase of $Y_{GGC}$. To compare theory with observations we have used the    
database by S00, which also provides the number of AGB stars for each    
cluster. These empirical data confirm the negligible effect of [Fe/H] on    
$R_2$; the mean value for the 43 clusters by S00 is $R_2$=0.14$\pm$0.05.    
    
In our models we have treated the convective mixing during central     
He-burning by    
including semiconvection, following the prescriptions by Castellani et    
al.~(1985). We have suppressed the onset of the breathing pulses    
during the latest phases of central He-burning by imposing that the allowed     
extension of the convective core does not lead to an increase of the    
central He abundance from one model to the next one (Caputo et al.~1989).    
All our models have reached the thermal pulses phase along the AGB;    
from this moment on the evolution is so fast that neglecting the    
computation of the thermal pulses does not affect the theoretical value of $R_2$.    
Our computations provide $R_2=$0.12,     
in good agreement with the empirical result;    
this confirms the adequacy of our mixing treatment in the HB stellar    
cores. If breathing pulses are not inhibited, HB evolutionary    
times are longer, due to the ingestion of fresh He into the central    
convective region following the onset of the pulses. We obtain in this case    
$R_2\sim$0.08, in disagreement with observations (a similar    
conclusion was reached by Caputo et al.~1989 by comparing their models    
with data about the GGC M5).      
    
We have also experimented with an alternative procedure     
to inhibit the onset of the    
breathing pulses; following the suggestions by Dorman \& Rood~(1993)    
we have set to zero the gravitational term in the energy generation    
equation for the central stellar regions during the later stage of    
core He-burning. In this way the breathing pulses are also effectively    
inhibited (see the detailed discussion by Dorman \& Rood~1993),    
and we obtained a decrease of     
both AGB and HB evolutionary time by about 2\% with respect to    
the procedure followed in our reference models; this leaves the value    
of $R_2$ unchanged ($R_2\sim$0.12), and causes a systematic increase     
of $Y_{GGC}$ by $\sim$0.003.    
    
The error on the individual $Y_{GGC}$ values displayed in Table~1  
comes basically from the random error on the individual  
He-abundance determinations (i.e., from the random error on the 
individual $R$-parameter estimates). In order to give a best estimate 
for $Y_{GGC}$ including also the sources of systematic error 
described before (associated to uncertainties in the theoretical 
models) and the effect of the still uncertain [Fe/H] scale, 
we have resorted to a Monte Carlo technique briefly explained in the 
following. We have considered as reference values for 
$Y_{GGC}$, the ones determined by adopting the CG97 [Fe/H] scale, 
taking into account the corrections 
for the blue HB clusters (lines 5 and 7 of Table~1 for, respectively, 
the Z00 and S00 sample); we notice that in case of the Z00 
data we consider the subsample unaffected by the precise choice of   
the clusters' age. Starting from each of these two reference $Y_{GGC}$ 
we have generated a set of 10000 synthetic 
He-abundance values, by applying (through a Monte Carlo simulation) 
to each generated abundance value a set of random and systematic errors, 
according to a given probability distribution. 
In particular, random errors have been modeled according to a Gaussian 
distribution with mean value equal to the reference one, and $\sigma$ 
equal to the corresponding random error on $Y_{GGC}$ provided in Table~1. 
The systematic uncertainties due to the choice of the [Fe/H] scale 
(which causes a decrease of $Y_{GGC}$ by 0.003 with respect to the 
reference value), $^{12}$C$(\alpha,\gamma)^{16}$O reaction rate 
(variation by $\pm$0.008), and breathing pulses suppression technique 
(increase by 0.003) have been modeled using an uniform 
distribution spanning the appropriate range. 
 
The mean values for the two final synthetic distributions of He-abundances 
are $Y_{GGC}$=0.243$\pm$0.006 in case of the Z00 sample,   
and $Y_{GGC}$=0.244$\pm$0.006 for the S00 sample.   
These values are, as expected, in very good reciprocal agreement and moreover   
compare well with the primordial He-abundance    
$Y_p$=0.248$\pm$0.001 inferred from the CMB in conjunction with    
primordial nucleosynthesis computations.    
    
Another important result of our analysis is the fact that there    
is no statistically significant increase of $Y_{GGC}$ with [Fe/H], at    
least within the analyzed GGC samples. This bears considerable    
interests for studies about Galactic chemical evolution.     
As a test for the reliability of this result we have performed the 
following numerical experiment. We have considered 
the clusters of the Z00 sample and the ZW84 metallicity scale
(we obtain an analogous result when using the CG97 scale); 
for each cluster we have considered a reference $R$ value obtained from our 
theoretical models, using a primordial He-mass fraction of 
0.245 and assuming a value for the chemical enrichment ratio     
$\Delta Y/\Delta Z$. We have then generated, using a Monte Carlo 
procedure, 10000 synthetic He-abundances for each individual 
cluster and a given choice of $\Delta Y/\Delta Z$, using a 
Gaussian distribution with mean value equal to the reference 
theoretical $R$ value and $\sigma$    
equal to the actual random error on observed $R$ value. 
 
For each of these synthetic samples we have then    
tried to recover the input $\Delta Y/\Delta Z$ value;      
we concluded from this analysis that ratios $\Delta Y/\Delta Z >$1 
should have been    
unambiguously detected even taking into account the actual    
observational errors on the determination of $R$.    
    
\section{Summary \label{conclusions_sect}}    
    
Following recent precise determinations of the primordial He-abundance    
coming from CMB analyses and primordial nucleosynthesis computations,    
we have rederived the initial He-abundance for stars in two samples of    
GGCs (Z00 and S00), using the $R$-parameter as abundance indicator.     
We have employed theoretical models computed adopting new and    
more accurate determinations of the EOS for the stellar matter and of the    
crucial $^{12}$C$(\alpha,\gamma)^{16}$O reaction rate.    
Our models include semiconvection, while the breathing pulses are    
suppressed, in agreement with the observational constraints coming    
from the measurements of the $R_2$ parameter in the S00 sample.    
    
By taking into account the uncertainties in the observed individual     
$R$ value, as well as the uncertainties in the GGC metallicity scale,    
the $^{12}$C$(\alpha,\gamma)^{16}$O reaction rate and the method      
for the breathing pulses suppression, we obtain     
$Y_{GGC}$=0.243$\pm$0.006 in case    
of the Z00 sample, and $Y_{GGC}$=0.244$\pm$0.006 in case of the S00    
sample. These abundances are in good reciprocal agreement    
and fully consistent with     
$Y_p$=0.248$\pm$0.001 recently determined from CMB analyses and     
primordial nucleosynthesis computations.    
Within the S00 sample we find a statistically significant spread    
of the individual He-abundances.  
This spread in the He-abundances is not found in the     
Z00 sample, and we argue that it is due to the inhomogeneity of the    
observational database used by S00, as opposed to the homogeneously    
observed and reduced photometry employed by Z00.    
    
It is important to remark that none of the two samples show     
any statistically significant increase of $Y_{GGC}$ with     
the cluster [Fe/H], a fact that is relevant in the context of the    
chemical evolution of the Galaxy.

\acknowledgments{     
S.C. and M.S. gratefully acknowledge the hospitality of     
the Max-Planck-Institut f\"ur Astrophysik,     
where a large part of this work has been carried out.  A.W.I. gratefully    
acknowledges partial financial support from an operating grant to    
Don A. VandenBerg from the Natural Sciences and Engineering Research    
Council of Canada. S.C. has been supported by MURST (Cofin2002). 
We wish thank the referee, E. Sandquist, for useful remarks 
which helped to improve the presentation of the paper.}

\pagebreak        

      
\clearpage    
\begin{deluxetable}{llcclc}     
\tablewidth{0pt}     
\tablecaption{Summary of $Y_{GGC}$ mean values and the associated 
intrinsic spread $\sigma_Y$, obtained by means of the F-test (see text 
for details).}     
\tablehead{     
\colhead{Sample}&      
\colhead{[Fe/H]}&      
\colhead{Blue HB correction}&     
\colhead{GGCs selection}&       
\colhead{$Y_{GGC}$}&       
\colhead{$\sigma_Y$}}     
  \startdata     
Z00 & CG97  & no  & $[Fe/H] < -$1.15 & 0.245$\pm$0.003 & 0.0\\    
    &       &     &   or $[Fe/H]> -$0.85  & & \\    
Z00 & ZW84  & no  &        all      & 0.242$\pm$0.003 & 0.0\\     
S00 & CG97  & no  &        all      & 0.246$\pm$0.005 & 0.03\\     
S00 & ZW84  & no  &        all      & 0.241$\pm$0.004 & 0.03\\     
Z00 & CG97  & yes & $[Fe/H] < -$1.15 & 0.243$\pm$0.003 & 0.0\\     
    &       &     &   or $[Fe/H]> -$0.85  & &  \\    
Z00 & ZW84  & yes &        all      & 0.240$\pm$0.003 & 0.0\\     
S00 & CG97  & yes &        all      & 0.244$\pm$0.004 & 0.03\\     
S00 & ZW84  & yes &        all      & 0.240$\pm$0.004 & 0.03\\     
\enddata     
\end{deluxetable}     
     
\clearpage    
    
      
\clearpage      
    
\clearpage      
    
\clearpage      
    
\end{document}